\documentclass[aps,pra,reprint,floatfix,superscriptaddress]{revtex4-1}

\usepackage{color}
\usepackage{graphicx}
\usepackage{amsmath}
\usepackage{amsfonts}

\newcommand{\lla}{\langle\!\langle}
\newcommand{\rra}{\rangle\!\rangle}

\begin{document}
\title{Entanglement generation through local field and quantum dissipation}
\date{\today}
\author{J\"urgen T. Stockburger}
\email{juergen.stockburger@uni-ulm.de}
\altaffiliation[current affiliation: ]{Institut f\"{u}r Komplexe
  Quantensysteme, Universit\"{a}t Ulm,
  Albert Einstein-Allee 11, 89069 Ulm,
  Germany}
\affiliation{Institut f\"{u}r Theoretische Physik, Universit\"{a}t Ulm,
  Albert Einstein-Allee 11, 89069 Ulm,
  Germany}
\affiliation{Center for Integrated Quantum Science and Technology,
  Albert Einstein-Allee 11, 89069 Ulm,
  Germany}

\author{Rebecca Schmidt}
\altaffiliation[current affiliation: ]{School of Mathematical Sciences,
The University of Nottingham, Nottingham, NG7 2RD, United Kingdom}
\affiliation{Institut f\"{u}r Theoretische Physik, Universit\"{a}t Ulm,
  Albert Einstein-Allee 11, 89069 Ulm,
  Germany}

\author{Joachim Ankerhold}
\altaffiliation[current affiliation: ]{Institut f\"{u}r Komplexe
  Quantensysteme, Universit\"{a}t Ulm,
  Albert Einstein-Allee 11, 89069 Ulm,
  Germany}
\affiliation{Institut f\"{u}r Theoretische Physik, Universit\"{a}t Ulm,
  Albert Einstein-Allee 11, 89069 Ulm,
  Germany}
\affiliation{Center for Integrated Quantum Science and Technology,
  Albert Einstein-Allee 11, 89069 Ulm,
  Germany}

\begin{abstract}
  Entanglement in a Gaussian two-mode system can be generated by local
  driving if additional non-local features are introduced to the
  dynamics. We demonstrate that weak to moderate ohmic friction
  arising from a dissipative environment can enable entanglement
  generation in a driven system. This synergy of driving and
  dissipation is highly sensitive to the pulse shape; several simple
  pulse shapes fail to produce this effect at all or deposit large
  amounts of energy in the system as a side effect. Complex pulse
  shapes, determined by optimal control techniques, however, are
  effective without detrimental side effects.
\end{abstract}

\maketitle
\section{Introduction}

Entanglement is a key resource in virtually all proposed applications
of quantum information and quantum computation. It is frequently
needed in the initial state of quantum algorithms. It must be provided
on a large scale in the ancilla degrees of freedom in most proposed
schemes for quantum error correction, and it is indispensable to
ensure privacy in quantum encryption.

There are many different technologies aiming for the realization of
the first universal quantum computer; among them several solid-state
approaches have been appreciated for some time due to their use of
highly scalable technologies. More recently, significant progress has
been made in significantly extending the decoherence times of
solid-state devices~\cite{bylan11,riget12}.

The dissipation mechanisms found in solid-state devices, in particular
circuit damping, correspond closely to the generic Ohmic damping
model~\cite{weiss12}. Here we study the effects of this dissipation
model in a minimal mechanism of entanglement generation between two
harmonic degrees of freedom. We find that the dissipative reservoir,
by itself a source of decoherence and entanglement decay, can be
turned into a \emph{source} of entanglement when combined with
\emph{local} external driving of the constituents of the entangled
pair which is being formed. Optimal control theory is used to find
pulse shapes which yield significant entanglement in a setting where
the only non-local interaction is quantum friction mediated by the
environment.

We compare our optimal control results to several simple guesses for
driving fields which might similarly induce entanglement. Of the three
alternative driving scenarios considered, two fail to produce
entanglement. One further scenario investigated results in
entanglement; however, this comes at the price of significantly
heating the system.

\section{Quantum information in continuous-variable systems}
\subsection{Gaussian quantum information}
Most practical proposals and algorithms in quantum information refer
to the concept of a qubit, i.e., the notion of two orthogonal quantum
states and their possible superpositions. The fundamental concepts of
quantum information, however, do not require this constructive
approach, and the important concepts of entanglement and non-classical
correlations equally apply to quantum systems with arbitrary level
structure, and can be defined without referring to a specific basis in
Hilbert space. In particular, Gaussian states of harmonic or
near-harmonic systems have recently attracted significant attention in
this context~\cite{weedb12,braun05}. In spite of their apparent simplicity and
near-classical appearance, these states can be viable sources of
entanglement. In this paper, we demonstrate a mechanism of
entanglement generation based on simple ingredients; control of local
fields and a common, Ohmic (featureless) thermal reservoir~\cite{weiss12}.

In the context of quantum optics, Gaussian states with non-negative
Wigner distribution are often labeled classical; however, when more
than one degree of freedom is considered, such states that look
``classical'' in their single-particle properties may display highly
non-classical features such as two-mode squeezing and
entanglement~\cite{galve10}. Several models have recently been
proposed for the generation or stabilization of these quantum
resources through suitably chosen couplings and/or external
fields~\cite{paz08,oliva11,horha08,galve10,seraf10}. Here we explore a
scenario so simple that it could almost be called a circumvention of
the well-known theorem that entanglement cannot be generated by a LOCC
mechanism (local operations, classical
communication)~\cite{schmi13}. In essence, we propose to entangle two
harmonic degrees of freedom, each weakly coupled to the same
dissipative reservoir, without any direct coupling to each other, by
applying suitable classical fields locally. In this paper, we
investigate the sensitivity of this mode of entanglement generation to
particular choices for the time dependence of external fields and their
potential side effects.

\subsection{Dynamics, decoherence and relaxation}

The type of entanglement generation we envisage is a dynamical
process, where the only non-local feature is the common dissipative
environment. Strong coupling to a dissipative reservoir has previously
been demonstrated to promote or statically support entanglement in
particular cases~\cite{horha08,mcend13}. Most of such approaches rely
on the simple fact that a strong dissipative mechanism will force a
quantum system into a corresponding pointer state~\cite{zurek81}. In a
somewhat more subtle approach, there is a decoherence-free subspace
instead of effective projections on individual pointer
states~\cite{horha08}. For these entangled states to be useful, they
need to be decoupled or transferred away from the reservoir after
preparation, with high fidelity required for this secondary
process. Here we do not rely on such a controlled switching of
dissipation, nor on any reliable transfer mechanism, but consider weak
to moderate coupling to the reservoir from the outset.

This means that the corresponding equilibrated two-mode state will
typically be separable. A thermal reservoir can also display effects
far more subtle than merely ``pulling towards'' pointer states. This
holds in particular for driven systems, which allow the ``mutual
friction'' of the two modes induced by the reservoir to be leveraged
for entanglement generation.

Quantum dynamics under the combined influence of strong driving and a
dissipative reservoir is a challenging theoretical subject even in the
case of weak dissipative coupling. In the absence of strong driving,
the reduced dynamics of the system is easily described by a
Lindblad-type master equation in most cases. However, since Lindblad
operators represent transitions between time-independent,
\emph{unperturbed} energy eigenstates of the system, this approach
fails to reproduce the true dynamics even qualitatively if strong
driving is taken into account by merely changing Hamiltonian part of
the Liouvillian, while keeping the Lindblad dissipator
unchanged~\cite{schmi11}.

An alternative approach to open system dynamics, which keeps external
driving and dissipation conceptually separate, relies on
Feynman-Vernon influence functionals~\cite{feynm63,weiss12}, which
have been used extensively in studying mesoscopic quantum
phenomena~\cite{makhl01}.
Feynman-Vernon influence functionals are Gaussian functionals of the
integrand in a path integral. Due to this property, even influence
functionals representing a \emph{quantum} reservoir can be constructed
from Gaussian noise which is \emph{classical} in the sense that it is
represented by $c$-numbers, but has correlation functions which
match the quantum case. Equivalently, they can be seen as
representing averages of sample states $\rho_{\xi}$ propagated under a
fictitious stochastic force~\cite{stock98,stock01,stock02}\footnote{In
  the classical limit, this stochastic force can be identified as
  thermal noise.}.

In the case of a simple Ohmic reservoir, the dynamics of sample states
for a single particle, given by the Feynman-Vernon path integral,
translates into a stochastic Liouville equation with dissipation
(SLED) \cite{stock99},
\begin{equation}
\frac{d}{dt} \rho_{\xi} = \frac{1}{i\hbar}
 \left( [H_{\mathrm{S}}, \rho_{\xi}] -
{\xi}(t) [q, \rho_{\xi}] \right)
 + \frac{\gamma}{2i\hbar} [q,\{p, \rho_{\xi}\}].
\label{eq:sled}
\end{equation}
Here $\xi(t)$ is a real-valued stochastic force with a spectrum
determined by the quantum fluctuation-dissipation theorem. In general,
this spectrum is non-white, and the corresponding time correlation
function decays only on the finite thermal time scale
$\hbar\beta$. One might interpret the last term in (\ref{eq:sled}) as
velocity-dependent friction; however, the partial term
$\{p,\rho_{\xi}\}$ does not represent objective or certain information
about the momentum. In the classical limit, on the other hand, this
interpretation is quite obvious, and (\ref{eq:sled}) is closely
related to the Klein-Kramers equation. A phase-space representation of
$\rho_{\xi}$ and an average over $\xi(t)$ (which turns into white
noise for $\hbar\to 0$) are all that is needed to make this
connection.

The key advantage of our approach in the context of \emph{driven}
systems lies in the fact that neither $\xi(t)$ nor the friction term
change when the system Hamiltonian $H_S$ is modified by external
driving. For two-mode Gaussian states and a quadratic Hamiltonian
\begin{equation}
H_{\rm S} = \sum_{k=A,B}\frac{p_k^2}{2
m}+\frac{m \omega^{2}}{2}q_k^{2}+\frac{u(t)}{2}q_k^{2}
\end{equation}
with parametric driving $u(t)$, (\ref{eq:sled}) can be mapped to a
set of ordinary differential equations for the parameters of the
Gaussian state. Here we choose the first and second cumulants of
position $q$ and momentum $p$ as parameters. In the simple case of two
independent reservoirs, the damping term $(\gamma/ 2i\hbar) [q,\{p,
\rho_{\xi}\}]$ is duplicated like the Hamiltonian, and we obtain the
set of equations ($k=A,B$)
\begin{align}
\label{eq:cumuB}
& \textstyle{\frac{d}{dt}} \lla q_k\rra = \lla p_k\rra/m\\
& \textstyle{\frac{d}{dt}} \lla p_k\rra = - (m\omega^2+u(t)) \lla q_k\rra -
\gamma \lla p_k\rra + \xi_k(t)\\
& \textstyle{\frac{d}{dt}} \lla q_k^2\rra = 2 \lla p_kq_k\rra/m\\
& \textstyle{\frac{d}{dt}} \lla p_k^2\rra = - 2\gamma\lla p^2_k\rra
-2m\omega^2 \lla p_kq_k\rra\\
& \textstyle{\frac{d}{dt}} \lla p_kq_k\rra = - m\omega^2 \lla q^2_k\rra
+ \lla p_k^2\rra/m - \gamma \lla p_kq_k\rra.
\label{eq:cumuE}
\end{align}
Symmetric (Weyl) operator ordering is assumed for all mixed products.

In the more interesting case of a shared reservoir coupling equally to
both modes, $\xi(t)$ loses its index $k$, and the new damping term
$(\gamma/ 2i\hbar) [q_A + q_B,\{p_A + p_B, \rho_{\xi}\}]$ in
(\ref{eq:sled}) is not itself a sum, but depends on sums of position
and momentum coordinates. Equations
(\ref{eq:cumuB})---(\ref{eq:cumuE}) must then be modified by A--B
cross terms proportional to $\gamma$ and augmented by additional
equations of motion for cumulants denoting correlations between modes
A and B. We obtain an expanded system of equations~\cite{schmi13a} of
the form
\begin{equation}
\label{eq:eomvec}
\dot{\vec{x}} = \mathsf{M}(u(t))\cdot\vec{x} + \xi(t) \vec{c},
\end{equation}
with
\begin{align}
\vec{x} = (
&
\lla q_A\rra, \lla p_A \rra, \lla q_B \rra, \lla p_B \rra,\nonumber\\
&\lla q^2_A\rra, \lla p^2_A \rra, \lla p_Aq_A \rra,
\lla q^2_B\rra, \lla p^2_B \rra, \lla p_Bq_B \rra,\nonumber\\
&\lla q_A q_B\rra, \lla p_A p_B \rra, \lla p_A q_B\rra, \lla p_B q_A\rra
)^{\dagger},
\end{align}
\begin{equation}
\mathsf{M} = \left(
\begin{array}{c|c}
\mathsf{M}_1 & 0\\
\hline
0 & \mathsf{M}_2
\end{array}
\right),
\end{equation}
\begin{equation}
\mathsf{M}_1 =
\left( \begin{array}{cccc}
0 & 1 & 0 & 0\\
-1\!-\!u & -\gamma &  0 & 0\\
0 & 0 & 0 & 1\\
0 & 0 & -1\!-\!u & -\gamma
\end{array}\right),
\end{equation}
\begin{equation}
\mathsf{M}_2 =
\left(
\begin{array}{cccccccccc}
0& 0& 2& 0& 0& 0& 0& 0& 0& 0\\
0& -2\gamma& 2a& 0& 0& 0& 0& -2\gamma& 0& 0\\
a& 1& -\gamma& 0& 0& 0& 0& 0& 0& -\gamma\\
0& 0& 0& 0& 0& 2& 0& 0& 0& 0\\
0& 0& 0& 0& -2\gamma& 2a& 0& -2\gamma& 0& 0\\
0& 0& 0& a& 1& -\gamma& 0& 0& -\gamma& 0\\
0& 0& 0& 0& 0& 0& 0& 0& 1& 1\\
0& -\gamma& 0& 0& -\gamma& 0& 0& -2\gamma& a& a\\
0& 0& 0& 0& 0& -\gamma& a& 1& -\gamma& 0\\
0& 0& -\gamma& 0& 0& 0& a& 1& 0& -\gamma
\end{array}
\right),
\end{equation}
$a = -1-u$ and
\begin{equation}
\vec{c} =
(0,1,0,1,0,0,0,0,0,0,0,0,0,0,0,0,0,0,0,0)^{\dagger}.
\end{equation}
For the sake of brevity, these expressions for $\mathsf{M}$ have been
given here using natural units of position and momentum (i.e., $m=1$ and
$\omega=1$).

The diagonalization of $\mathsf{M}$ is time-dependent, with the
parametric driving $u(t)$ yet unknown. Equation (\ref{eq:eomvec}) must
therefore be solved numerically for any given $u(t)$.

\section{External control and optimization of entanglement generation}
\subsection{Optimal control theory\label{sec:OCT}}
Freedom to choose the time dependence of the field $u(t)$, combined
with the intent to maximize the entanglement of the final two-mode
state, defines an elaborate optimization
problem~\cite{steng86}. Maximization of the final-state entanglement
is subject to a dynamical constraint given by (\ref{eq:eomvec}).
This constraint is valid at any time, and, in our case, for any
realization of the stochastic process $\xi(t)$. Lagrange multipliers
$\vec{\lambda}(t)$ referring to these constraints are therefore (i)
functions of time and (ii) random variables in the same probability
space as $\xi(t)$.  Variations $\delta \vec{x}(t)$ are independent for
different realizations of the noise $\xi(t)$. The control field
$u(t)$, on the other hand, is not a random variable, since the
optimization objective is based on the final quantum state, resulting
from an average over realizations.

Having introduced the dynamics through Lagrange multipliers, the
optimization objective is now the sum of the entanglement measure we
want to maximize and a Lagrange multiplier term in the form of a time
integral of a suitably chosen expectation value in the probability
space of the stochastic process $\xi(t)$,
\begin{equation}
  \int_0^{t_f}\!\!dt\, 
  \mathbb{E}\left[-(\vec\lambda, \dot{\vec{x}})
    + \mathcal{H}(\vec{x},\vec{\lambda},u,\xi)\right]
\end{equation}
with
\begin{equation}
\mathcal{H}(\vec{x},\vec{\lambda},u,\xi) = \left(\vec{\lambda}, 
\mathsf{M}(u)\cdot\vec{x} +
  \xi\,\vec{c}\right).
\end{equation}
Round parentheses denote the scalar product. Variational calculus
determines the time dependence of the Lagrange multipliers to be
governed by the equation of motion
\begin{equation}
\label{eq:eomlambda}
\dot{\vec{\lambda}} = - \mathsf{M}^{\dagger}(u) \vec{\lambda},
\end{equation}
with a boundary condition of the form $\vec{\lambda}_f = g(\vec{x}_f)$
at the \emph{end time}. The function $g$ is determined by the
variation of the final-state entanglement measure with respect to
$\delta \vec{x}_f$. Since the Lagrange multipliers $\vec{\lambda}$ are
dynamical variables, they are frequently referred to as
\emph{co-states}~\cite{steng86}. Even though (\ref{eq:eomlambda}) is a
deterministic equation of motion, the co-states are true random
variables since their boundary condition is random.
Equations (\ref{eq:eomvec}) and (\ref{eq:eomlambda}), together with
\begin{equation}
\label{eq:gradH}
\frac{\partial}{\partial u}
 \mathbb{E}\left[ \mathcal{H}(\vec{x},\vec{\lambda},u,\xi)\right]= 0,
\end{equation}
again valid at arbitrary time $t$, constitute necessary conditions for
a local minimum of the constrained optimization problem. The
approach of constructing co-states based on the ordinary differential
equations (\ref{eq:eomvec}) provides significant advantages
over the direct application of control theory to the two-particle
version of the Liouville equation
(\ref{eq:sled})~\footnote{J. T. Stockburger, unpublished}.

Due to the mixed boundary conditions, the simultaneous solution of
(\ref{eq:eomvec}), (\ref{eq:eomlambda}) and (\ref{eq:gradH}) is
usually computed by iteration, starting with an initial guess for
$u(t)$, propagating first $\vec{x}$, then $\vec{\lambda}$, finally
interpreting $\partial\, \mathbb{E}\left[ \mathcal{H} \right]\!/\partial
u$ as the gradient associated with the constrained optimization
problem. This provides a search direction for an updated test function
$u(t)$, which may be used, e.g., in a gradient search.

Here we do not use a simple gradient search, but the iterative
algorithm of Krotov~\cite{kroto83,konno99,sklar02}, where the co-state
dynamics has a slightly different role. The typical advantages of
Krotov's algorithm are decreased sensitivity to the initial guess,
sizable improvements during the first few iterations, and monotonicity
of the iteration without additional algorithmic elements such as line
searches.

The entanglement measure we use is logarithmic
negativity~\cite{vidal02,weedb12}, which adequately quantifies
entanglement for Gaussian states. In the present case, it can easily
be computed from the covariance matrix $\sigma$ associated with the
observables $q_A$, $p_A$, $q_B$ and $p_B$. The logarithmic negativity
can be obtained from the determinants of $\sigma$ and its $2\times 2$
submatrices,
\begin{equation}
\sigma = \left(
\begin{array}{c|c}
\alpha & \delta \\
\hline
\delta^t & \beta
\end{array}
\right),
\end{equation}
as
\begin{equation}
E_{\mathcal{N}} = \max\{0, -\ln \nu_-\}
\end{equation}
with
\begin{equation}
\nu_- = \sqrt{2} \sqrt{|\alpha| + |\beta| - 2 |\delta| - \sqrt{\mu}}
\end{equation}
and
\begin{equation}
\mu = (|\alpha| + |\beta| - 2 |\delta|)^2 - 4 |\sigma|.
\end{equation}

In the discussion of some of our results, we omit the maximum function
and present plots of $-\ln \nu_-$ instead of the logarithmic
negativity. Negative values of $-\ln \nu_-$ give some indication how
far a separable state is from nearby entangled states.

\begin{figure}
\includegraphics[width=\columnwidth]{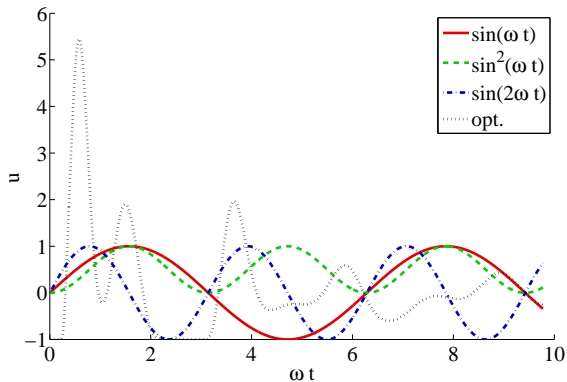}
\caption{\label{fig:sig} Control signals to be compared: iteratively
  optimized pulse~\cite{schmi13}, dotted/red, and periodic signals as
  indicated in the inset. The control field $u(t)$ is scaled by its
  natural unit $m\omega^2$.}
\end{figure}

\subsection{Numerical results}

In a recent publication~\cite{schmi13}, we have demonstrated the
feasibility of entanglement generation by local control and a shared
weak dissipative reservoir. Neither of these factors alone can induce
entanglement; an entangled state results only if dissipation is
\emph{combined} with a properly shaped pulse. We observe that
entanglement generation coincides with a build-up of two-mode
squeezing. In a perfectly symmetric setting, the two-mode squeezing
can be visualized as simple squeezing of the symmetric and
antisymmetric normal modes~\footnote{These modes are degenerate in the
  absence of a static coupling.}. However, two-mode squeezing is not
sufficient for entanglement generation. The Hamiltonian control term
\begin{equation}
H_{\rm c}(u) = \textstyle{\frac{u^2}{2}} \left(q_A^2 + q_B^2\right)
= \textstyle{\frac{u^2}{2}} \left(q_{+}^2 + q_{-}^2 \right)
\end{equation}
is clearly a local operator, therefore the external driving described
by it does not generate entanglement. Without dissipation, the
symmetric and antisymmetric modes (coordinates $q_{+}$ and $q_{-}$)
are squeezed equally. With a common dissipative reservoir, coupling
only to $q_{+}$, the symmetric and antisymmetric modes are
no longer treated on an equal footing. Dissipation then curbs the
build-up of squeezing in the symmetric mode (but not the antisymmetric
mode). This effect, and the attendant generation of entanglement can
be promoted by particular pulse shapes $u(t)$.

\begin{figure}
\includegraphics[width=\columnwidth]{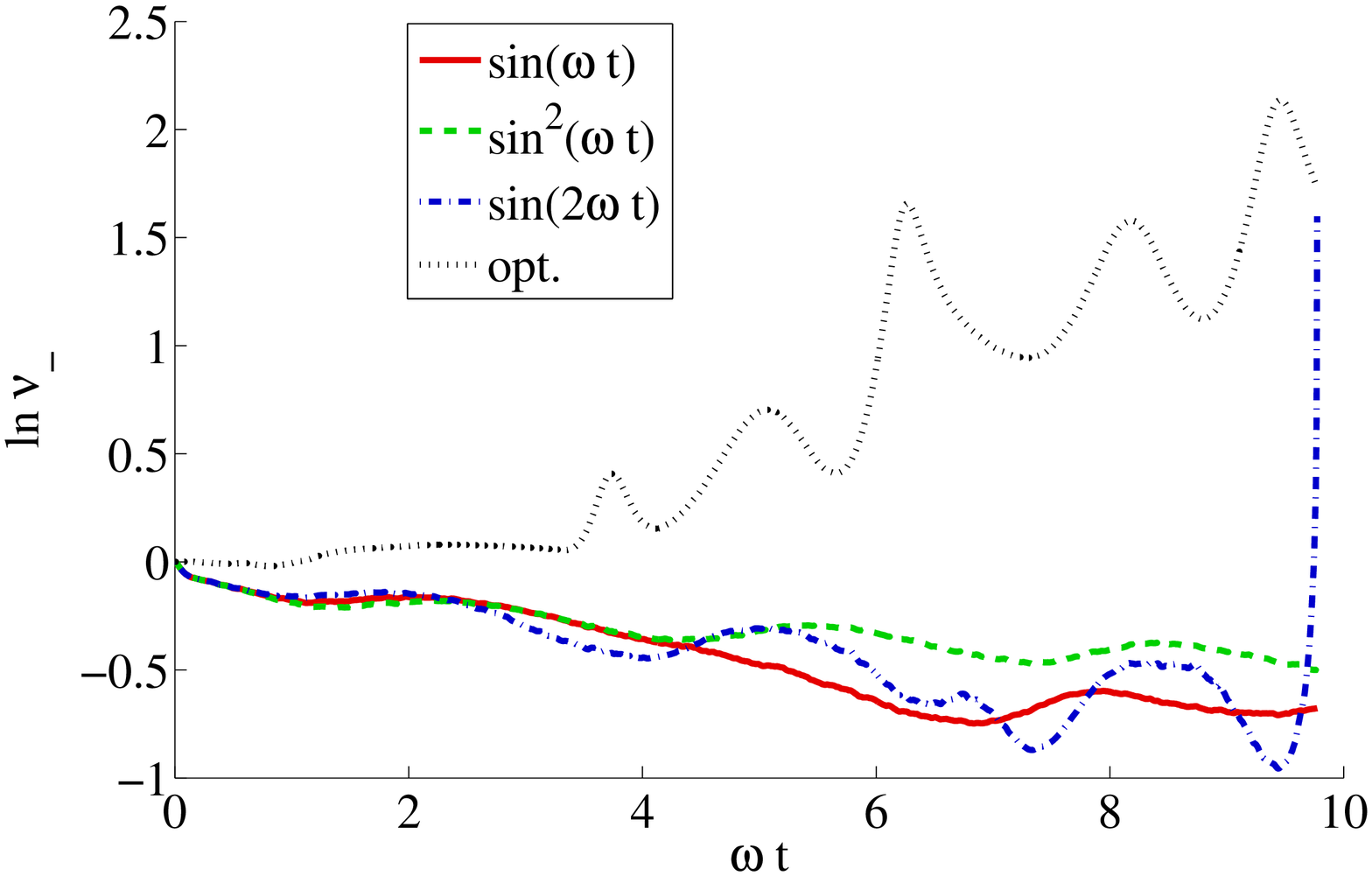}
\caption{\label{fig:neg}
Entanglement generation for different pulse shapes $u(t)$ (see
figure~\ref{fig:sig}). Note that the curves indicate an
entanglement measure only for positive values---negative
values give some indication of ``how far'' a separable state is from
nearby entangled states, see section \ref{sec:OCT}.}
\end{figure}

\begin{figure}
\includegraphics[width=\columnwidth]{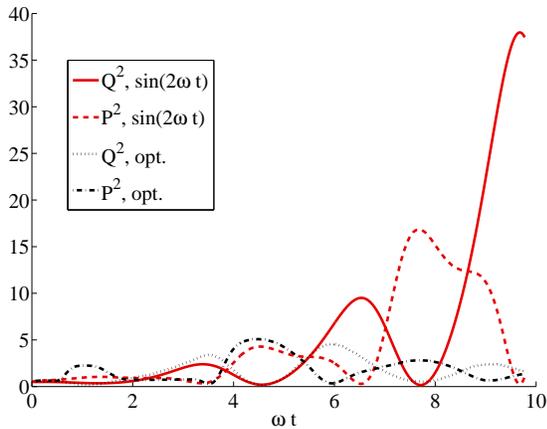}
\caption{\label{fig:q2p2}
Second moments of position and momentum, indicating potential and
kinetic energy for different pulse shapes $u(t)$ (see
figure~\ref{fig:sig}).}
\end{figure}

Optimized pulse shapes $u(t)$ have been determined by repeated Krotov
iterations. They typically display a complex structure (see
figure~\ref{fig:sig}, dotted curve). It is not a priori obvious
whether this structure is incidental, maybe due to a near degeneracy
of the optimization problem, or if it is necessary and cannot be
substituted for by simple pulse shapes. In order to address this
question, we compare the optimized solution with the effect of simple
periodic pulse shapes (i) $u(t) = \sin(\omega t)$, (ii) $u(t) =
\sin^2(\omega t)$, and (iii) $u(t) = \sin(2 \omega t)$.

In cases (i) and (iii), $u(t)$ briefly touches the value $-1$,
transforming the oscillatory degree of freedom into a free particle
for an instant. Cases (ii) and (iii) are examples of driving at
frequency $2\omega$, which amounts to resonant driving for the case of
parametric control. The resulting dynamics is illustrated by
figure~\ref{fig:neg}, indicating the amount of entanglement achieved,
and figure~\ref{fig:q2p2}, indicating how much kinetic and potential
energy is accumulated by driving the system. The reservoir temperature
has been chosen as $T = \hbar\omega/k_{\rm B}$; damping is moderate,
$\gamma/\omega = 0.1$.

Out of the three simple pulse shapes examined, only case (iii) creates
a final state which is entangled. Figure~\ref{fig:q2p2} shows that this
simplicity comes at a price: Parametric driving at the first harmonic
tends to add huge amounts of energy to the modes, a side effect that
should be undesirable in most applications.

Our data indicates that the numerical optimization provides
substantial benefits over several simple versions of an ansatz form
the pulse shape: Optimizing entanglement rather than guessing pulse
shapes yields benefits even for figures of merit which were not
explicitly included as objectives of the computation.

\section{Conclusions and outlook}

Numerical techniques of optimal control are a useful tool in quantum
information. We have introduced a simple model with features that do
not directly promote entanglement (weak dissipation, local driving),
but do not contradict it, either (non-locality from the shared
environment). Even though this system tends to equilibrate towards a
separable state, its dissipative features can be leveraged by
optimized parametric driving which provides two-mode squeezing, which
is further modified by dissipation to yield entanglement. Pulse shapes
found by numerical optimization are superior to the simple ansatz of
periodic functions at the natural frequency of the system or its
double. One might ask whether the shared reservoir could have the same
effect if it has a significant spatial extension. If so, this would
allow the entanglement of spatially separated modes \textit{in
  situ}. Ohmic reservoirs in the parameter range considered here are
typical of solid-state quantum devices, suggesting an experimental
realization with existing resources.

\section*{Acknowledgments}
This work was supported by Deutsche Forschungsgemeinschaft
through grants AN336/6-1 and SFB/TRR21.

\bibliographystyle{unsrt}
\bibliography{fqmt13proc}

\begin{thebibliography}{10}

\bibitem{bylan11}
Jonas Bylander, Simon Gustavsson, Fei Yan, Fumiki Yoshihara, Khalil Harrabi,
  George Fitch, David~G. Cory, Yasunobu Nakamura, Jaw-Shen Tsai, and William~D.
  Oliver.
\newblock Noise spectroscopy through dynamical decoupling with a
  superconducting flux qubit.
\newblock {\em Nat. Phys.}, 7(7):565--570, July 2011.

\bibitem{riget12}
Chad Rigetti, Jay~M. Gambetta, Stefano Poletto, B.~L.~T. Plourde, Jerry~M.
  Chow, A.~D. C\'orcoles, John~A. Smolin, Seth~T. Merkel, J.~R. Rozen,
  George~A. Keefe, Mary~B. Rothwell, Mark~B. Ketchen, and M.~Steffen.
\newblock Superconducting qubit in a waveguide cavity with a coherence time
  approaching 0.1 ms.
\newblock {\em Phys. Rev. B}, 86:100506, Sep 2012.

\bibitem{weiss12}
Ulrich Weiss.
\newblock {\em Quantum dissipative systems}.
\newblock World Scientific, Hackensack, NJ, 4th edition, 2012.

\bibitem{weedb12}
Christian Weedbrook, Stefano Pirandola, Ra\'ul Garc\'ia-Patr\'on, Nicolas~J.
  Cerf, Timothy~C. Ralph, Jeffrey~H. Shapiro, and Seth Lloyd.
\newblock {Gaussian} quantum information.
\newblock {\em Rev. Mod. Phys.}, 84:621--669, May 2012.

\bibitem{braun05}
Samuel~L. Braunstein and Peter van Loock.
\newblock Quantum information with continuous variables.
\newblock {\em Rev. Mod. Phys.}, 77:513--577, Jun 2005.

\bibitem{galve10}
Fernando Galve, Leonardo~A. Pach\'on, and David Zueco.
\newblock Bringing entanglement to the high temperature limit.
\newblock {\em Phys. Rev. Lett.}, 105:180501, Oct 2010.

\bibitem{paz08}
Juan~Pablo Paz and Augusto~J. Roncaglia.
\newblock Dynamics of the entanglement between two oscillators in the same
  environment.
\newblock {\em Phys. Rev. Lett.}, 100:220401, Jun 2008.

\bibitem{oliva11}
Stefano Olivares and Matteo G.~A. Paris.
\newblock Fidelity matters: The birth of entanglement in the mixing of gaussian
  states.
\newblock {\em Phys. Rev. Lett.}, 107:170505, Oct 2011.

\bibitem{horha08}
Christian H\"orhammer and Helmut B\"uttner.
\newblock Environment-induced two-mode entanglement in quantum {Brownian}
  motion.
\newblock {\em Phys. Rev. A}, 77:042305, Apr 2008.

\bibitem{seraf10}
Alessio Serafini and Stefano Mancini.
\newblock Determination of maximal {Gaussian} entanglement achievable by
  feedback-controlled dynamics.
\newblock {\em Phys. Rev. Lett.}, 104:220501, Jun 2010.

\bibitem{schmi13}
Rebecca Schmidt, J\"urgen~T. Stockburger, and Joachim Ankerhold.
\newblock Almost local generation of {Einstein}-{Podolsky}-{Rosen} entanglement
  in nonequilibrium open systems.
\newblock {\em Phys. Rev. A}, 88:052321, Nov 2013.

\bibitem{mcend13}
S.~McEndoo, P.~Haikka, G.~De Chiara, G.~M. Palma, and S.~Maniscalco.
\newblock Entanglement control via reservoir engineering in ultracold atomic
  gases.
\newblock {\em EPL (Europhysics Letters)}, 101(6):60005, 2013.

\bibitem{zurek81}
W.~H. Zurek.
\newblock Pointer basis of quantum apparatus: {I}nto what mixture does the wave
  packet collapse?
\newblock {\em Phys. Rev. D}, 24:1516--1525, 1981.

\bibitem{schmi11}
R.~Schmidt, A.~Negretti, J.~Ankerhold, T.~Calarco, and J.~T. Stockburger.
\newblock Optimal control of open quantum systems: Cooperative effects of
  driving and dissipation.
\newblock {\em Phys. Rev. Lett.}, 107:130404, Sep 2011.

\bibitem{feynm63}
R.~P. Feynman and F.~L. Vernon.
\newblock The theory of a general quantum system interacting with a linear
  dissipative system.
\newblock {\em Ann. Phys. (N.Y.)}, 24:118, 1963.

\bibitem{makhl01}
Yuriy Makhlin, Gerd Sch\"on, and Alexander Shnirman.
\newblock Quantum-state engineering with {J}osephson-junction devices.
\newblock {\em Rev. Mod. Phys.}, 73:357--400, 2001.

\bibitem{stock98}
J.~T. Stockburger and C.~H. Mak.
\newblock Dynamical simulation of current fluctuations in a dissipative
  two-state system.
\newblock {\em Phys. Rev. Lett.}, 80:2657--2661, 1998.

\bibitem{stock01}
J.~T. Stockburger and H.~Grabert.
\newblock Non-{M}arkovian quantum state diffusion.
\newblock {\em Chem. Phys.}, 268:249--256, 2001.

\bibitem{stock02}
J.~T. Stockburger and H.~Grabert.
\newblock Exact $c$-number representation of non-{M}arkovian quantum
  dissipation.
\newblock {\em Phys. Rev. Lett.}, 88:170407, 2002.

\bibitem{Note1}
In the classical limit, this stochastic force can be identified as thermal
  noise.

\bibitem{stock99}
J.~T. Stockburger and C.~H. Mak.
\newblock A stochastic {L}iouvillian algorithm to simulate dissipative quantum
  dynamics with arbitrary precision.
\newblock {\em J. Chem. Phys.}, 110:4983--4985, 1999.

\bibitem{schmi13a}
Rebecca Schmidt.
\newblock {\em Cooperative phenomena in open quantum systems subject to
  external control}.
\newblock PhD thesis, Universit\"at Ulm, 2013.

\bibitem{steng86}
Robert~F. Stengel.
\newblock {\em Stochastic optimal control: theory and application}.
\newblock A Wiley-Interscience publication. Wiley, New York, NY, 1986.

\bibitem{Note2}
J. T. Stockburger, unpublished.

\bibitem{kroto83}
V.~F. Krotov and I.~N. Fel'dman.
\newblock An iterative method for solving optimal control problems.
\newblock {\em Engineering cybernetics}, 17:123--130, 1983.

\bibitem{konno99}
A.~I. Konnov and V.~F. Krotov.
\newblock On global methods of successive improvement of controlled processes.
\newblock {\em Automation and Remote Control}, 60:1427--1436, 1999.

\bibitem{sklar02}
Shlomo~E. Sklarz and David~J. Tannor.
\newblock Loading a bose-einstein condensate onto an optical lattice: An
  application of optimal control theory to the nonlinear schr\"odinger
  equation.
\newblock {\em Phys. Rev. A}, 66:053619, 2002.

\bibitem{vidal02}
G.~Vidal and R.~F. Werner.
\newblock Computable measure of entanglement.
\newblock {\em Phys. Rev. A}, 65:032314, Feb 2002.

\bibitem{Note3}
These modes are degenerate in the absence of a static coupling.

\end{thebibliography}
\end{document}